# SENSATION: an Authoring Tool to support Event–State paradigm in End-User Development

Giuseppe Desolda[1], Francesco Greco[1], Francisco Guarnieri[2],
Nicole Mariz[2], Massimo Zancanaro[2,3]

[1]Computer Science Department, University of Bari Aldo Moro, Italy
[2]Department of Psychology and Cognitive Science, University of Trento, Italy
[3]Fondazione Bruno Kessler

```
giuseppe.desolda@uniba.it, francesco.greco@uniba.it
      francisco.guarnieri@studenti.unitn.it,
         nicole.mariz@studenti.unitn.it,
         massimo.zancanaro@unitn.it
```

```
   preprint version - please cite as
   Desolda G., Greco F., Guarnieri F.,
      Mariz N., Zancanaro M. (2021)
   SENSATION: An Authoring Tool to Support
      Event-State Paradigm in End-User
   Development. In: Ardito C. et al. (eds)
   Human-Computer Interaction - INTERACT
   2021. INTERACT 2021. Lecture Notes in
   Computer Science, vol 12933. Springer,
   Cham. https://doi.org/10.1007/978-3-
               030-85616-8_22
```

**Abstract.** In this paper, we present the design and the evaluation of an authoring tool for End-User Development, which supports the definition of Trigger-Actions rules that combines events and states in the triggers. The possibility of using either states or events in triggers has already been discussed in the literature. However, it is recognized that the state/event distinction is difficult to manage for users. In this paper, we propose an authoring tool that provides explicit support for managing this distinction. We compare it with a state-of-the-art authoring tool that implements the classical event-event paradigm.

## 1 Introduction

Trigger-Action programming (TAP) is emerging as the most adopted paradigm for supporting end-users, particularly those without IT skills, in defining the behavior of *Internet-of-Things* devices and digital web services. TAP is a simplified form of the *Event Condition Action* (ECA), a common approach for rule-based systems, originally employed to manage databases [14] and control industrial processes [18]. However, when applied in the form of *Trigger-Action* rules for End-User Development (EUD), the *Condition* part is usually left out for the sake of simplicity, and the rules take the simple form of "*IF <a trigger occurs> THEN <an action is executed>*". This kind of rule has become popular on web services like IFTTT [16] and Zapier [29]. More advanced solutions propose the possibility to trigger a rule when different events co-occur or execute a chain of actions [10]. Some tools allow the specification of a condition but usually as part of a generic "IF" trigger, and it is usually not represented in the syntactic form of the rule (for example, [10, 14, 22]).

As noted by Brackenbury and colleagues [4], a source of confusion in the TAP paradigm derives from the fact that triggers indicate both instantaneous events or state, and users are not always able to understand the difference between the two [15, 28], causing inconsistencies, loops, and redundancies in the behaviors of the smart objects [7].

The work presented in this paper builds upon and extends the approach proposed by Huang and Cakmak [15], which recognizes the need to differentiate these two types of triggers. It also integrates the findings from Gallitto and colleagues [13] that argue how using different language forms may help the users better understand the semantic and operationalization of a set of rules. We discuss the design and an initial evaluation of an authoring tool, SENSATION, for the EUD of rules in a constrained form. It explicitly requires a single event that triggers the rule and conjunction of states in which the world is expected to be for the rule to be executed as a specific form of condition in the ECA approach. The proposed tool structures and guides the construction of a rule by contextual filtering the available choices regarding events and states.

## 2      Background and related work

Internet-of-Things (IoT) has now been established as one of the most widely used technologies in recent years, and recent forecasts tell that it will become even more pervasive in the next years. In 2019, around 26 billion devices were installed worldwide and, by the end of 2025, over 75 billion devices are expected[1]. This proliferation of technology brings several challenges, ranging from technical aspects to hardware miniaturization, energy consumption, security, cost, and aspects related to interaction with smart devices.

One of the most critical challenges concerns the possibility for non-technical users to customize the behavior of these devices to better satisfy their specific needs [12]. Customization might be required to personalize a single application's behavior (like a thermostat) or connect several IoT objects, each designed to solve a specific task, with the purpose to realize a more complex combined service. The need to create these combinations among devices is growing as the availability of smart objects increases.

---

[1] https://financesonline.com/iot-trends/

Several Task Automation Systems (TAS) [9] have been proposed as web-based tools that allow users to compose smart objects' behavior through visual interfaces that support the TAP paradigm. Among the most popular there are IFTTT [16] and Zapier [29], EFESTO-5W [10], EFESTO-4SIE, [1, 2], Microsoft Flow [23], Mozilla's Things Gateway [24], SmartRules [26], NinjaBlocks [17].

Although the possibility of exploiting TAP for End-User Development has been widely demonstrated [5, 14, 15, 21], several critical aspects have been noted in the literature on the possibility for the users to understand and manage the potential complexity of rules. In this respect, explicit support for debugging has been proposed [7, 21, 25], and (semi-)automatic methods to prevent undesired effects have been experimented with [8]. Concerning other tools for rules with explicit conditions, a notable exception might be the tool proposed by Troung and colleagues [27], which allows the syntactical specification of the location ("WHERE") in which the event should take place in order for the action to be executed. Similarly, EFESTO-5W supports the specification of both temporal (WHEN) and spatial (WHERE) conditions in the trigger. Even if these TASs have much success due to their simplicity, many limitations result in ECA rule errors like inconsistencies, loops, and redundancies [7].

The need to differentiate events from states in TAP has been recognized as a source of possible misunderstanding by users. Brackenbury and colleagues [4]] present ten programming bugs grouped in 1) bugs in control flow, 2) timing-related bugs, and 3) errors in user interpretation. They stated that one of the most important causes is the temporal aspect of triggers and actions [15, 28] since users are often confused when they had to distinguish triggers based on events (i.e., that occur in a specific moment in time) and states (i.e., that are true over a time span). This research focuses on the temporal aspect of triggers in TAP, proposing and evaluating a tool that provides explicit support for state/event distinction.

## 3 The SENSATION tool

SENSATION (Smart EveNt and State AcTION rules) is a tool for the EUD of Trigger-Action rules designed to facilitate the distinction between events and state. Our starting point was that this distinction might be challenging to articulate by users but essential to recognize and reduce at least specific common bugs [4, 13-15].

Our first design assumption is that a constrained syntactic structure may help the user recognize the different roles of events and states in a rule [13, 15]. In the interface, this is implemented by structuring the rules on the form "DO <action> WHEN <event> WHILE <states specifications>".

The users might find it challenging to articulate the distinction between events and states [14, 15] (for example, between the event "the door is opening" and the state "the door is open"). Therefore, our second design assumption is that a filtering approach may alleviate the need to make such a distinction. In this respect, our interface structures the filling of the three parts of a rule by providing filtered access to actions, events, and state specifications. **Fig. 1.** displays the main screen of the SENSATION interface. It has two principal areas: the top area that displays the three main parts of a rule, namely the action (called DO), the trigger event (called WHEN), and the state (called WHILE); and the bottom area in which the elements that can be chosen according to the specific part of the rule are listed.

SENSATION proposes a constrained process to construct triggering rules by which the user has to answer three questions: 1) what will be done? 2) what happens? 3) in which configuration of states should hold for the rule to trigger? The two final questions are meant to help the user to manage the distinction between events and states correctly.

Although the interface implicitly suggests starting from the action part, the user can decide whether to start from the event or the states. While one of the parts is selected, the specific elements are listed in the bottom area. The DO part can be filled with one action or a sequence of actions. The WHEN part has to be filled with exactly one event. The WHILE is optional; if present, it contains one or more state specifications. A video reporting an example of a rule created with SENSATION is available at this link https://youtu.be/_TOuFC8ghgI.

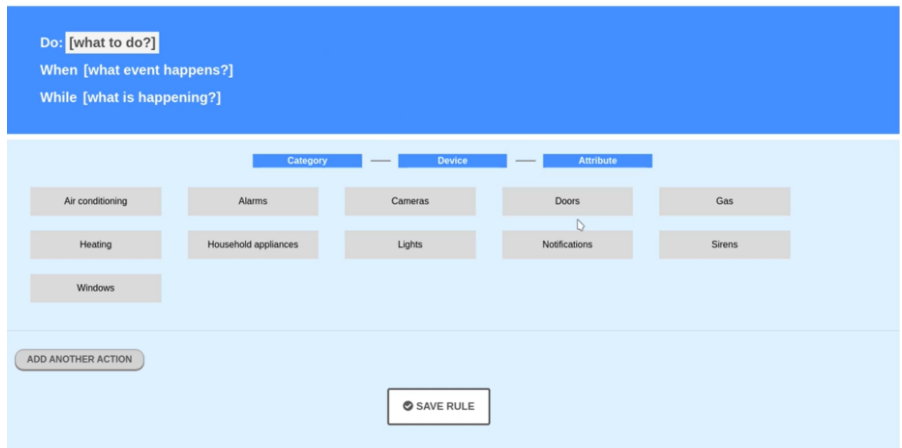

**Fig. 1.** The main screen of SENSATION where the users start the creation of rules by composing the DO (action), WHEN (event) and WHILE (the states). In the screen, the DO part is selected and the actual devices are listed: once selected the device, the tool will propose to select the specific actions available for that specific device.

## 4 The user study

The study has been designed as a *within-subject* study with SENSATION as the experimental condition and EFESTO-5W [1, 2, 10] as a control condition. The study's main objective was to assess if the explicit support for choosing events and states in SENSATION improves accuracy in writing Trigger Action Rules with respect to the state-of-the-art TAP approaches.

From the SENSATION design assumptions, we posed the following hypothesis for the experimental study:

*H1: SENSATION induces a more accurate definition of rules than EFESTO-5W, for those tasks in which the distinction between events and states is crucial (as discussed above);*

*H2: the time to complete successful tasks with SENSATION should not be longer than with EFESTO-5W, despite it more strongly constrains the interaction.*

*H3: the perceived usability of SENSATION is not lower than the usability of EFESTO-5W, despite it more strongly constrains the interaction.*

### 4.1 Materials

EFESTO-5W [10] has been selected as a control condition because *i)* it outperforms popular tools like IFTTT [16]; *ii)* it can be customized in term of smart devices and digital services to be used in the ECA rules; and iii) it supports multiple events, states, and actions but without explicitly support the distinction between events and states.

Both SENSATION and EFESTO-5W have been configured with the same descriptions of smart devices, events and actions for a scenario of smart home control.

For the study, four tasks have been created. The first task requires a simple event/action rule: it has been used to provide a baseline between the two systems (we did not expect a difference in performance with this task). The other three tasks are based on the classes of bugs for TAP proposed by Brackenbury and colleagues [4]. The tasks have been proposed in the form of scenarios for which participants had to write the rules, (in both SENSATION and EFESTO-5W) to realize them (see **Table 1**).

**Table 1.** List of tasks in the form of scenarios and their rationale in the study

| # | Task scenarios | Rationale for the tasks |
|---|---|---|
| T1 | You want to have cameras record who is buzzing your home. | Simple task that requires only one event and no states, it is used as a baseline to compare the systems |
| T2 | It starts to rain but you are not at home (Via Roma 2, Milan) and you want to make sure that no water gets in through the windows. | Medium difficulty task that may induce a *time window fallacy* bug |
| T3 | With the alarm on, there must be no open windows. | Difficult task that may induce a *flipped triggers* bug |
| T4 | The camera detects that you are approaching the door of your house (via Roma 2, Milan) and you want the door to open automatically. | Difficult task that may induce a *contradictory triggers* bug |

### 4.2 Measures

For what concerns quantitative data, measures of performance and measures of perceived usability have been collected. The performance has been assessed by manually annotating the correct (referred to as "S" for success below) and the incorrect (referred to as "F" for failure) task execution. Furthermore, a task has been classified as partially corrected (referred to as "P") in those cases in which additional spurious elements are added (like more actions than requested or redundant states). Task time completion has also been recorded for successful and partially successful tasks. The SUS (System Usability Scale) questionnaire [6] has been used for measuring perceived usability, and the UMUX-Lite questionnaire [20] for user experience. Regarding qualitative data, the errors in the tasks have been analyzed and classified.

### 4.3    Participants and procedure

Due to COVID-19 restrictions, the study had to be performed remotely. The tool *eGLU-Box PA* [11] has been used to perform the remote study. The participants have been recruited among the students at the University of Bari and at the University of Trento on a voluntary basis.

Forty (40) volunteers (14 females) were recruited. Their mean age was 21.7 years (SD = 3.46). All of them declared a good knowledge of IT technology (7.8/10, SD=1.28), a medium experience with programming languages (4.8/10, SD=4.81), a good experience with the use of IoT technology (6.2/10, SD=2.0), and medium knowledge of tools for IoT configuration (5.1/10, SD=2.28).

The participants received an email with a detailed description of the study and the specification of the required software and hardware (PC, microphone, a standard browser, and a stable internet connection), the link to the pre-questionnaire, and the link to start the test in eGLU-Box PA. The participants were free to decide when to do the study, but they were asked to start within few days from the reception of the email and complete it in one round.

The conditions and tasks were randomized among participants following a Latin-square design to avoid the carry-over effect. After the task execution, eGLU-Box PA administered the SUS [6] and UMUX-Lite questionnaires [20].

### 4.4    Results

Two participants experienced technical problems, and they are not included in the analysis, which therefore considers 38 participants.

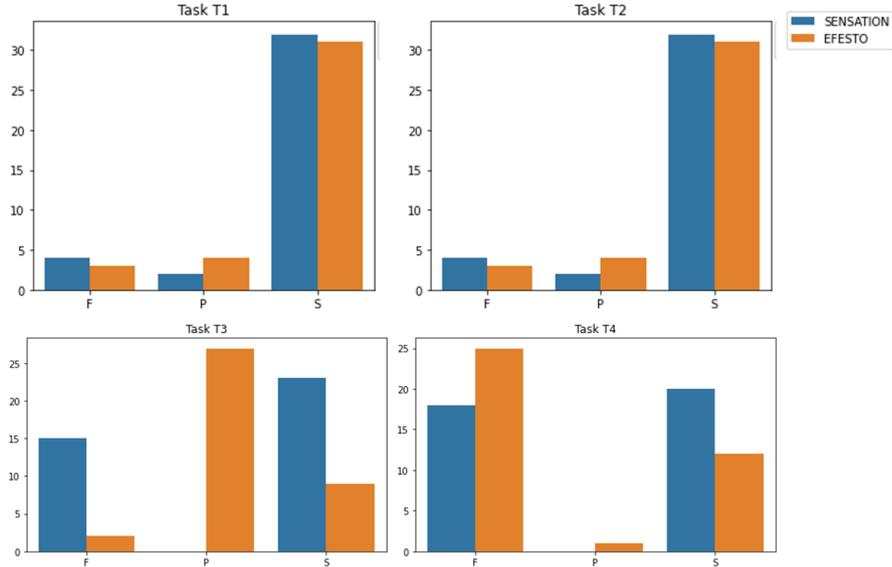

**Fig. 2.** Histograms depicting the distribution of successes (S), partial successes (P) and failures (F) for the 4 tasks in the two tools.

**Analysis of performance – task success (H1). Fig. 2.** shows the distribution of successes (S), partial successes (P), and failures (F) for the 4 tasks in the two systems. SENSATION seems to perform better than EFESTO in all tasks but the first one. In order to analyze the differences, we assigned a score of 2 to each successful task, a score of 1 to each partial successful task, and 0 to each failed task. Since the distributions are not normally distributed, we used the non-parametric Wilcoxon test, which also accounts for repeated measures.

Overall, SENSATION has a better performance rate with an average score of 1.19 (SD=0.89) while EFESTO has an average score of 0.98 (SD=.97); the difference is statistically significant Wilcoxon w=550.0, p<0.01).

Significant differences in the mean scores are found for task T2 (SENSATION x̄=0.7, SD=0.94; EFESTO x̄=0.34, SD= 0.75; Wilcoxon w=22.5, p<0.05) and task T4 (SENSATION x̄=1.05, std=1.10, EFESTO x̄=0.66, std=0.94; Wilcoxon w=6.0, p<0.05).

The differences in task T1 and T3 are not statistically significant (for T1 SENSATION x̄=1.74, SD=.64, EFESTO x̄=1.74, SD=0.60; Wilcoxon w=10.5, p=1.0; for T3: SENSATION x̄=1.2, SD=0.99, EFESTO x̄=1.18, SD=0.51; Wilcoxon w=211.0, p=0.876).

**Qualitative analysis of errors (H1).** In total, 130 errors have been detected and analyzed. As expected, the errors are not distributed evenly across tasks and participants. Task T1 had a minimal number of errors in both systems. These errors are usually related to the wrong choice of the event or the action. The most error-prone tasks are T2 (with 58 errors in total, 22 in SENSATION and 36 in EFESTO) and T4 (45 errors in total, 20 in SENSATION, and 25 in EFESTO). That is not surprising since both these tasks require a distinction between an event (respectively, the ringing of the bell and the detection of movement by the camera) and a state (the person's location). Task T3 is more complicated. This task required understanding the alarm both as an event and as a state. In EFESTO, the users made fewer errors (2 vs. 15 in SENSATION) but there a lot of partial successes (27 vs no partial successes in SENSATION) and fewer successes (9 vs. 23 in SENSATION). The most common type of error in SENSATION was the addition of unrelated events. In few cases, the participants confused the event with the action. In EFESTO, participants chose the wrong event or the wrong action. The partial successes were mainly due to the choice of either the event or the state related to the alarm but not both. Furthermore, this task has been perceived as ambiguous by many users. As a check, we tried the statistical tests on the performances described above without considering task T3. The results are comparable with a significant difference overall and for T2 and T4, and no significant difference for T1.

**Analysis of performance – task time (H2).** For what concerns the time to completion (for successful and partially successful) tasks, SENSATION has an average time of 82 seconds (SD=38.97) and EFESTO an average time of 97 seconds (SD=52.49). Yet, there is a high variability (from a minimum of 20 seconds to a maximum of over 5 minutes: x̄=89.5, SD=52.49). Furthermore, the distributions are not normally distributed, and the variances are not equal. At the same time, the two samples are not independent. Therefore, none of the standard tests can be applied. Kruskal-Wallis test suggests that the means are not statistically different (w=2.7, p=0.097), while Welch test suggests a difference w=-2.13, p<0.05 (it is worth noting that the Kruskal-Wallis' condition of independence is violated as well as the Welch's condition of normality).

**Analysis of perceived usability (H3).** The SUS scores highlighted that SENSATION had a good usability level with an average score of 70.6/100 (SD= 18.1) [3]. The high scores are also maintained in the two SUS subscales [19]: Usability has a mean score of 78.3 SD=16.4) and Learnability has a mean score of 80.9 (SD=16.3).

EFESTO-5W, too, demonstrated high scores with a mean global SUS score of 77.2/100 (SD=15.1); for the Usability subscale, the mean score is 72.2/100 (SD=17.3) and for the Learnability scale, the mean score is 70.1/100 (SD= 22.2). Paired sample t-test revealed that there are no statistical significant differences between the two systems in term of SUS and its subscales ($t_{SUS\_GLOBAL}(37)$=1.819, p=0.077; $t_{SUS\_USABILITY}(37)$=1.689, p=0.099; $t_{SUS\_LEARNABILITY}(37)$=1.729, p=0.092).

Similarly, the analysis of UMUX-Lite results revealed overall good UX of both EFESTO (x̄ = 5.6, SD= 1.2) and SENSATION (x̄ = 5.3, SD= 1.2), and even in this case, no differences emerged applying the paired sample t-test (t(37)=1.108, p=0.274).

## 5     Discussion and conclusion

The study results indicate that the tasks have generally performed well both in EFESTO and in SENSATION. The simpler task T1, used a baseline, had a very high success score for both tools. The tasks T2 and T4, whose correct definition depended on an accurate distinction between events and states, had a better performance in SENSATION, as hypothesized (H1). That may suggest that the structured approach offered by SENSATION is effective in this respect. For time of execution, the statistical analysis may weakly suggest that SENSATION allowed a quicker completion of the tasks despite it more strongly constrains the interaction, as hypothesized (H2). Finally, both systems have a similar high score on usability, as hypothesized (H3).

Overall, we can claim that SENSATION provides effective support in managing the distinction between events and state without complicating the whole approach.

The study described has several limitations. In particular, task T3 raised some interpretation problems by users that make it difficult to properly analyze the performance. Although the results seemed to be robust even without considering it, this might impact the analysis of the users' experience. Furthermore, participants were just briefly exposed to both systems with minimal training. Therefore, the tasks had been kept simple. Longer and ecological studies with more meaningful tasks are needed.

As future work, we planned to refine SENSATION taking into account the limitations that emerged in this study. The new version of the tool will be also evaluated by performing a wider and deeper controlled experiment with more users and in different domains, as well as during a longitudinal study in real contexts.

## Acknowledgment

This work is partially supported by the Italian Ministry of University and Research (MIUR) under grant PRIN 2017 "EMPATHY: EMpowering People in deAling with internet of THings ecosYstems.".

## References

1. Ardito, C., Desolda, G., Lanzilotti, R., Malizia, A. and Matera, M. (2019). Analysing Trade-offs in Frameworks for the Design of Smart Environments. *Behaviour & Information Technology*, 39(1), 47-71.
2. Ardito, C., Desolda, G., Lanzilotti, R., Malizia, A., Matera, M., Buono, P. and Piccinno, A. (2020). User-defined semantics for the design of IoT systems enabling smart interactive experiences. *Personal and Ubiquitous Computing*, 24(6), 781-796.
3. Bangor, A., Kortum, P. and Miller, J. (2008). The system usability scale (SUS): An empirical evaluation. *International Journal of Human-Computer Interaction*, 24(6), 574-594.
4. Brackenbury, W., Deora, A., Ritchey, J., Vallee, J., He, W., Wang, G., Littman, M.L. and Ur, B. (2019). How Users Interpret Bugs in Trigger-Action Programming. In *Proc. of the Human Factors in Computing Systems (CHI '19)*. Association for Computing Machinery, Paper 552.
5. Brich, J., Walch, M., Rietzler, M., Weber, M. and Schaub, F. (2017). Exploring End User Programming Needs in Home Automation. *ACM Transaction on Computer-Human Interaction*, 24(2), Article 11 (May 2017), 1-35.
6. Brooke, J. (1996). SUS-A quick and dirty usability scale. *Usability evaluation in industry*, 189(194), 4-7.

7. Corno, F., Russis, L.D. and Roffarello, A.M. (2019). Empowering End Users in Debugging Trigger-Action Rules. In *Proc. of the Conference on Human Factors in Computing Systems (CHI '19)*. Association for Computing Machinery, Paper 388.

8. Corno, F., Russis, L.D. and Roffarello, A.M. (2020). TAPrec: supporting the composition of trigger-action rules through dynamic recommendations. In *Proc. of the International Conference on Intelligent User Interfaces (IUI '20)*. Association for Computing Machinery, 579–588.

9. Coronado, M. and Iglesias, C.A. (2016). Task Automation Services: Automation for the Masses. *IEEE Internet Computing*, 20(1), 52-58.

10. Desolda, G., Ardito, C. and Matera, M. (2017). Empowering end users to customize their smart environments: model, composition paradigms and domain-specific tools. *ACM Transactions on Computer-Human Interaction*, 24(2), Article 12 (April 2017), 52 pages.

11. Federici, S., Mele, M.L., Lanzilotti, R., Desolda, G., Bracalenti, M., Buttafuoco, A., Gaudino, G., Cocco, A., Amendola, M. and Simonetti, E. (2019). Heuristic Evaluation of eGLU-Box: A Semi-automatic Usability Evaluation Tool for Public Administrations. In *Proc. of the International Conference on Human-Computer Interaction (HCII '19)*. Springer International Publishing, 75-86.

12. Fischer, G., Giaccardi, E., Ye, Y., Sutcliffe, A. and Mehandjiev, N. (2004). Meta-design: a manifesto for end-user development. *Communications of the ACM*, 47(9), 33-37.

13. Gallitto, G., Treccani, B. and Zancanaro, M. (2020). If when is better than if (and while might help): on the importance of influencing mental models in EUD (a pilot study). In *Proc. of the 1st International Workshop on Empowering People in Dealing with Internet of Things Ecosystems - co-located with International Conference on Advanced Visual Interfaces (AVI 2020) (EMPATHY '20)*. CEUR-WS.

14. Ghiani, G., Manca, M., Paternò, F. and Santoro, C. (2017). Personalization of Context-Dependent Applications Through Trigger-Action Rules. *ACM Transaction on Computer-Human Interaction*, 24(2), Article 14 (April 2017), 33 pages.

15. Huang, J. and Cakmak, M. (2015). Supporting mental model accuracy in trigger-action programming. In *Proc. of the ACM International Joint Conference on Pervasive and Ubiquitous Computing*. Osaka, Japan, Association for Computing Machinery,215–225.

16. IFTTT Inc. *IFTTT*. Retrieved from https://ifttt.com/ Last Access June 1, 2019.

17. Inc., N.B. *Ninja Blocks*. Retrieved from https://github.com/ninjablocks Last Access April 10, 2021.

18. Joonsoo, B., Hyerim, B., Suk-Ho, K. and Yeongho, K. (2004). Automatic control of workflow processes using ECA rules. *IEEE Transactions on Knowledge and Data Engineering*, 16(8), 1010-1023.

19. Lewis, J. and Sauro, J. (2009). The Factor Structure of the System Usability Scale. In: M. Kurosu Ed. *Human Centered Design - HCD 2009*. LNCS, Vol. 5619, Springer Berlin Heidelberg, 94-103.

20. Lewis, J.R., Utesch, B.S. and Maher, D.E. (2013). UMUX-LITE: when there's no time for the SUS. In *Proc. of the Conference on Human Factors in Computing Systems (CHI '13)*. ACM, New York, NY, USA, 2099-2102.

21. Liang, C.-J.M., Bu, L., Li, Z., Zhang, J., Han, S., Karlsson, B.F., Zhang, D. and Zhao, F. (2016). Systematically Debugging IoT Control System Correctness for Building Automation. In *Proc. of (BuildSys '16)*. Association for Computing Machinery, 133–142.

22. Metaxas, G. and Markopoulos, P. (2017). Natural Contextual Reasoning for End Users. *ACM Transaction on Computer-Human Interaction*, 24(2), Article 13 (May 2017), 1-36.

23. Microsoft. *Microsoft Flow*. Retrieved from https://flow.microsoft.com/ Last Access February 28th, 2021.

24. Mozilla. *WebThings Gateway* Retrieved from https://iot.mozilla.org/gateway/ Last Access April 10, 2021.

25. Russis, L.D. and Roffarello, A.M. (2018). A Debugging Approach for Trigger-Action Programming. In *Proc. of the Extended Abstracts of the 2018 CHI Conference on Human Factors in Computing Systems*. Montreal QC, Canada,   Association for Computing Machinery,Paper LBW105.

26. SmartThings. *SmartRules*. Retrieved from https://smartrulesapp.com/ Last Access April 10, 2021.

27. Truong, K.N., Huang, E.M. and Abowd, G.D. (2004). CAMP: A Magnetic Poetry Interface for End-User Programming of Capture Applications for the Home. In *Proc. of the International Conference on Ubiquitous Computing (UbiComp '04)*. Springer Berlin Heidelberg, 143-160.

28. Ur, B., Ho, M.P.Y., Brawner, S., Lee, J., Mennicken, S., Picard, N., Schulze, D. and Littman, M.L. (2016). Trigger-Action Programming in the Wild: An Analysis of 200,000 IFTTT Recipes. In *Proc. of the SIGCHI Conference on Human Factors in Computing Systems (CHI '16)*. ACM, New York, NY, USA, 3227-3231.

29. Zapier Inc. *Zapier*. Retrieved from https://zapier.com/ Last Access May 9, 2021.